# Broadband angle- and permittivity-insensitive nondispersive optical activity based on chiral metamaterials

Kun Song,[1] Min Wang,[1] Zhaoxian Su,[1] Changlin Ding,[1] Yahong Liu,[1] Chunrong Luo,[1]

Xiaopeng Zhao,[1] Khagendra Bhattarai,[2] & Jiangfeng Zhou[2]

[1]Department of Applied Physics, Northwestern Polytechnical University, Xi'an, 710129, China.

[2]Department of Physics, University of South Florida, 4202 East Fowler Ave, Tampa, FL, 33620-5700. Correspondence and requests for materials should be addressed to K.S. (email: songkun@nwpu.edu.cn) or J.Z. (email: jiangfengz@usf.edu)

Because of the strong inherent resonances, the giant optical activity obtained via chiral metamaterials generally suffers from high dispersion, which has been a big stumbling block to broadband applications. In this paper, we propose a type of chiral metamaterial consisting of interconnected metal helix structures with four-fold symmetry, which exhibits nonresonant Drude-like response and can therefore avoid the highly dispersive optical activity resulting from resonances. It shows that the well-designed chiral metamaterial can achieve nondispersive and pure optical activity with high transmittance in a broadband frequency range. And the optical activity of multi-layer chiral metamaterials is proportional to the layer numbers of single-layer chiral metamaterial. Most remarkably, the broadband behaviors of nondispersive optical activity and high transmission are insensitive to the incident angles of electromagnetic waves and permittivity of dielectric substrate, thereby enabling more flexibility in polarization manipulation.

**Introduction**

Ever since a long time ago, manipulating the polarization states of electromagnetic waves has been of great scientific interest due to the important applications in realms of life sciences, photoelectrons, telecommunications, *etc*. Materials with chirality, which can rotate the polarization plane of electromagnetic waves, are very suitable for designing polarization converters. Nevertheless, since the chirality in natural materials is weak and dispersive, the ordinary polarization converters made of natural materials usually suffer from huge thicknesses being much larger than the operating wavelengths or narrow operating bandwidths, severely impeding the practical applications in the low-frequency region and micro-nano devices. Thus, materials that possess strong chirality are highly desired.

Metamaterials enable numerous of extraordinary electromagnetic phenomena that do not exist in natural materials, for instance, abnormal refraction or reflection[1-6], super-resolution imaging[7,8], cloaking[9-11], perfect absorption of electromagnetic waves[12-14]. The presence of metamaterials makes it possible for us to obtain strong chirality. In the past few years, chiral metamaterials (CMMs) have been constructed to realize negative refractive index[15-20], strong optical activity[21-25], circular dichroism[26-30], as well as asymmetric transmission[31-35]. Using strong resonances, the optical rotatory power of CMMs might rise up to several orders of magnitudes larger than that of natural substances[36-40]. However, the giant optical activity of the previous CMMs is generally accompanied by high losses, high dispersion, narrow transmission bandwidth, and even strong circular dichroism owing to the inherent Lorentz-like resonances[18-20,36,41-43], which is exceedingly detrimental for designing broadband and efficient polarization rotators. Recently, there has been much effort devoted to exploring how to achieve nondispersive optical activity. By combining a meta-atom with its complement in a chiral configuration, low dispersive optical activity at the transmission resonance has been demonstrated; while this kind of CMMs are subjected to narrow transmission bandwidths[44-46]. In some recent papers, three-dimensional off-resonant or nonresonant types of CMMs were also proposed to accomplish nondispersive optical activity[47,48]. However, the high technology and complex process needed for fabricating three-dimensional structures are still tremendous obstacles, especially at the optical part of the spectrum. Therefore, despite some promising progress, there remains a great need for the further development in CMMs with more simplicity of the fabrication and more applicability under multi-application situation.

In this paper, we propose an intriguing CMM that is composed of helix structures. Unlike the Lorentz-like resonances in the previous CMMs[18-20,36-38,41,42], the present CMM occurs Drude-like response on account of the interconnected metal structures, exhibiting nonresonant feature. The simulation, calculated, and experimental results show that the elaborate CMM can achieve strong nondispersive optical activity for a linearly polarized incident wave simultaneously along with high transmittance and extremely low ellipticity in a broadband frequency region. And the optical activity of multilayer CMMs is proportional to the number of CMM layers. It is worth noting that the broadband behaviors of nondispersive optical activity and high transmission of the single-layer CMM can be maintained independent of the incident angles of electromagnetic waves and the permittivity of dielectric substrate, which are unprecedented electromagnetic properties that have not been demonstrated by the CMMs reported previously[19-21,36-38,41,42]. Moreover, it is also found that the transmission frequency of the proposed CMM presents dynamical tunability by altering the permittivity of dielectric substrate, indicating that the CMM is very suitable for designing frequency-tunable polarization manipulation devices. Due to the excellent performances, the considered CMM holds great promises for the potential applications in realm such as telecommunications, among others.

**Results**

**Design of unit cells and theoretical calculation.** Figure 1(a,b) respectively show the schematic view and photograph of the present CMM, of which the unit cells are composed of double helix structures rotated in the four-fold (C4) symmetry along the propagating direction of electromagnetic waves, ensuring a pure optical activity effect. The metal slices on the front and back surfaces of the dielectric substrate are connected via metallization holes. Obviously, all the unit cells of the designed CMM are interconnected. The geometrical parameters of the unit cells are as follows: $a$ = 3 mm, $b$ = 1.3 mm, $d$ = 0.6 mm, $g$ = 0.3 mm, $l$ = 1.7 mm, $r$ = 1.5 mm, $t$ = 3.07 mm, and $w$ = 0.9 mm. The metal copper cladding is 0.035 mm in thickness with a conductivity of $\sigma$ = 5.8×10$^7$ S/m. The permittivity of the F4BM-2 substrate is 2.65+0.001*$i$. In the experiments, a CMM sample composed of 40*40 unit cells was fabricated via the printed circuit board etching process.

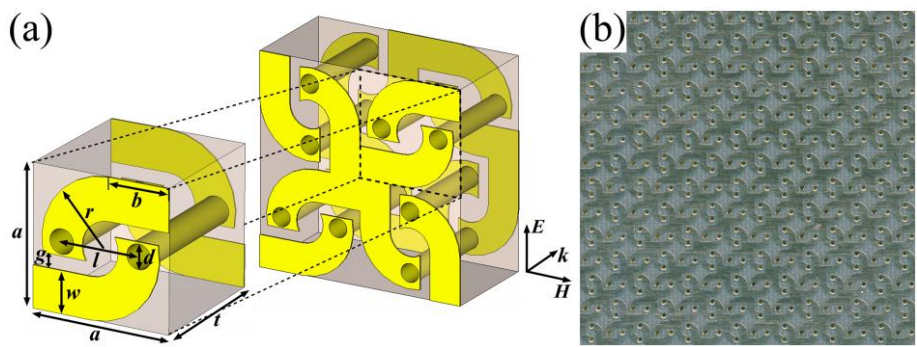

**Figure 1.** (**a**) Schematic diagram and (**b**) photograph of the proposed CMM.

To theoretically get the electromagnetic properties of the present CMM, we use the effective current model to calculate the effective parameters of the CMM[48]. Due to the interconnected metal structures, the current can flow freely in the CMM, which will lead to a Drude-like response. For the double helix structure, the inductance and resistance are approximatively calculated as $L = \mu_0 R(\ln\frac{8R}{w} - 2)$ and $R_\Omega = \sqrt{\frac{\mu_0 \omega S}{2\sigma w^2}}$, respectively, where $R = \sqrt{S} = \sqrt{(l+d)t}$, $S$ is the effective cross section area of the double helix structure[48-50]. In the case of time-harmonic electromagnetic field, the electric potential can be expressed as:

$$\xi = a E - (\pm i\omega \mu_0 S H) = (R_\Omega - i\omega L\, I). \qquad (1)$$

Here, the signs $\pm$ represent the right-handed and left-handed helix structure; $I$ is the total current which flows in the helix structure. From equation (1), we can obtain:

$$I = \frac{aE}{R_\Omega - i\omega L} - \frac{\pm i\omega \mu_0 S H}{R_\Omega - i\omega L}. \qquad (2)$$

Then, the electric and magnetic dipole moments can be obtained via the following formulas[48]:

$$\boldsymbol{P} = \frac{\boldsymbol{J}}{-i\omega} = \frac{a\boldsymbol{I}}{-i\omega V} = -\frac{a^2}{(i\alpha\omega + \omega^2)LV}\boldsymbol{E} - \frac{\pm\mu_0 aS}{(\alpha - i\omega)LV}\boldsymbol{H}, \qquad (3)$$

$$\boldsymbol{M} = \pm\frac{\boldsymbol{I}S}{V} = \frac{\pm aS}{(\alpha - i\omega)LV}\boldsymbol{E} - \frac{i\omega\mu_0 aS^2}{(\alpha - i\omega)LV}\boldsymbol{H}, \qquad (4)$$

where $V$ is the volume of a unit cell and $\alpha = \frac{R_\Omega}{L}$ is the dissipation constant. From equations (3) and (4), the electric displacement vector $\boldsymbol{D}$ and magnetic flux density vector $\boldsymbol{B}$ can be expressed as follows:

$$\boldsymbol{D} = \varepsilon_0 \boldsymbol{E} + \boldsymbol{P} = [\varepsilon_0 - \frac{a^2}{(i\alpha\omega + \omega^2)LV}]\boldsymbol{E} - \frac{\pm\mu_0 aS}{(\alpha - i\omega)LV}\boldsymbol{H}, \qquad (5)$$

$$\boldsymbol{B} = \mu_0(\boldsymbol{H} + \boldsymbol{M}) = \frac{\pm\mu_0 aS}{(\alpha - i\omega)LV}\boldsymbol{E} + \mu_0\boldsymbol{H} - \frac{i\omega\mu_0^2 aS^2}{(\alpha - i\omega)LV}\boldsymbol{H}. \qquad (6)$$

For the chiral media, the constitutive equation is as follows:

$$\begin{pmatrix} D \\ B \end{pmatrix} = \begin{pmatrix} \varepsilon_0 \varepsilon & -i\kappa c \\ i\kappa c & \mu_0 \mu \end{pmatrix} \begin{pmatrix} E \\ H \end{pmatrix}, \quad (7)$$

where $\kappa$ is the chirality parameter. According the equations (5)-(7), the effective parameters of the proposed CMM can be derived as follows:

$$\varepsilon = \varepsilon_f - \frac{a^2}{(\omega^2 + i\omega\alpha)\varepsilon_0 LV} \quad (8)$$

$$\mu = \mu_f - \frac{\omega \mu_0 S^2}{(\omega + i\alpha)LV} \quad (9)$$

$$\kappa = \pm \frac{\mu_0 caS}{(\omega + i\alpha)LV} \quad (10)$$

where $\varepsilon_f$ and $\mu_f$ are the quantitative fitting parameters. The transmission spectrum of the CMM is expressed in the terms of the following equation[48]:

$$T = \frac{4Z e^{ink_0 h}}{(1+Z)^2 - (1+Z)^2 e^{2ink_0 h}}. \quad (11)$$

Here, $Z = \sqrt{\mu/\varepsilon}$ is the effective impedance; $n = \sqrt{\mu\varepsilon}$ is the effective average refractive index; $k_0 = \omega\sqrt{\mu_0\varepsilon_0}$ is the wave number in vacuum. And the polarization azimuth rotation angle can be calculated as:

$$\theta = \frac{\mathrm{Re}(\kappa)\omega h}{c}. \quad (12)$$

**Results of single-layer CMM.** Figure 2 shows the simulated, calculated and measured results of the single-layer CMM at normal incidence. It is seen that the experimental results are in good qualitative agreement with the simulated and calculated ones. In Fig. 2(a,b), we illustrate the transmission spectra of the proposed CMM. It can be found that a transmission peak with the transmittance rising up to nearly unity occurs at 3.4 GHz. And the transmittance is over 0.8 in the range of 2.9 ~ 4.0 GHz with a relative bandwidth of 32%, indicating that the present CMM can achieve broadband and high-efficiency transmission. The results of the polarization azimuth rotation angle of the CMM are plotted in Fig. 2(c,d). It is obvious that the polarization azimuth rotation angle is kept about 45° within the entire measured frequency region, indicating that the CMM can realize broadband and nondispersive optical activity. Figure 2(e,f) depict the results of the ellipticity $\eta$. As we know, the chiral media will generate a pure optical activity effect at $\eta = 0$, *i.e.*, when a linearly polarized incident wave passes through the chiral media, the transmitted wave remains linearly polarized but with the polarization plane rotated by an angle of $\theta$[19]. Since the values of ellipticity in Fig. 2(e,f) are nearly zero in the whole frequency range, the transmission spectrum is still linearly polarized without distortion of polarization states.

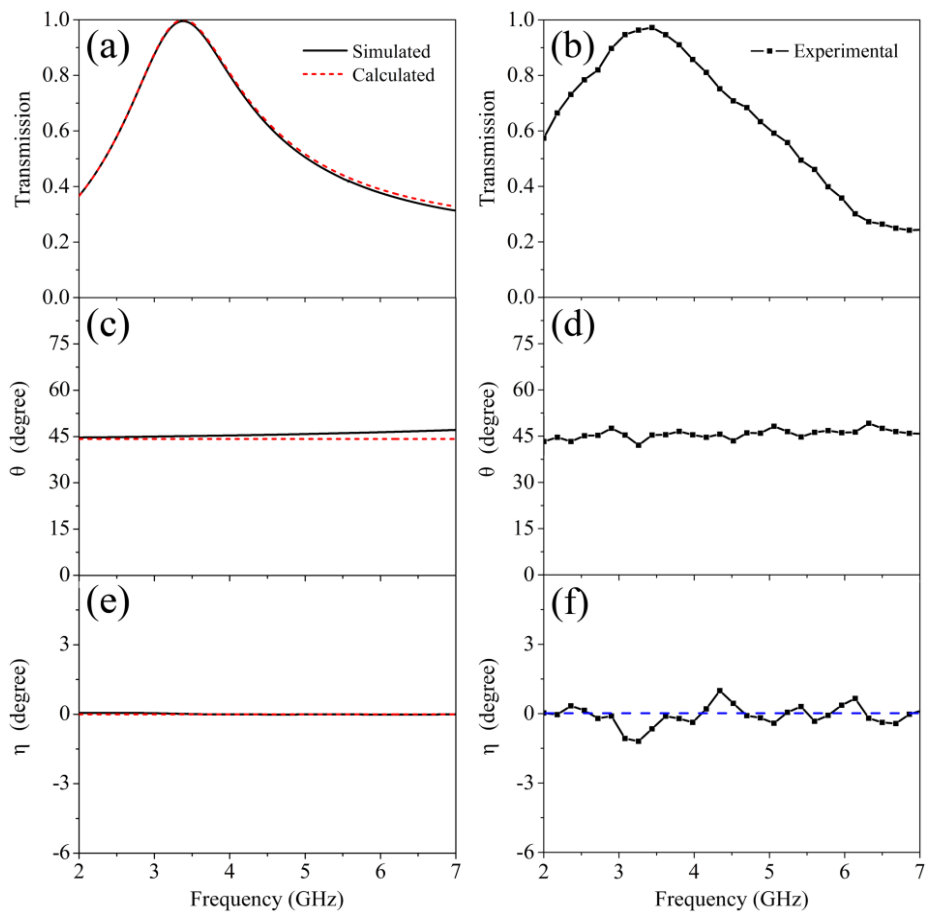

**Figure 2.** The simulated, calculated, and experimental results of the single-layer CMM at normal

incidence. (**a**) and (**b**) Transmission spectra ($T = \sqrt{T_{xx}^2 + T_{yx}^2}$), (**c**) and (**d**) Polarization azimuth rotation angle $\theta$, (**e**) and (**f**) Ellipticity $\eta$. In the calculation, the quantitative fitting parameters $\varepsilon_f$ and $\mu_f$ are chosen as 21.3 and 0.95, respectively.

The electromagnetic properties of the single-layer CMM at oblique incidence are shown in Figure 3. The incident angle of electromagnetic wave is tuned by a step of 10°. Figure 3(a,b) present the simulation and experimental transmission spectra evolving with different incident angles. It can be found that, although the transmission peak generates a very slight blue shift simultaneously companied by a slight reduction of the maximum transmittance as the incident angle increases, the broadband and efficient transmission performance still exists. In Fig. 3(c,d), it is of significance that the polarization azimuth rotation angle $\theta$ in the whole frequency region is kept approximately 45° with the incident angle increasing from 0° to 40°, which implies that the single-layer CMM can realize nondispersive optical activity in a wide region of incident angle. As shown in Fig. 3(e,f), the ellipticity of the single-layer CMM gradually increases with response to the increment of the incident angle. However, the maximum value of the ellipticity is less than 1.0° even if the incident angle rises up to 40°. As the ellipticity is very small, the transmission spectrum of the single-layer CMM can still be regarded as linearly polarized. These facts reveal that the broadband high transmission and nondispersive optical activity behaviors of the single-layer CMM can be maintained regardless of the incident angles, exhibiting more advantages than the CMMs previously reported[18-20,36-38,41,42].

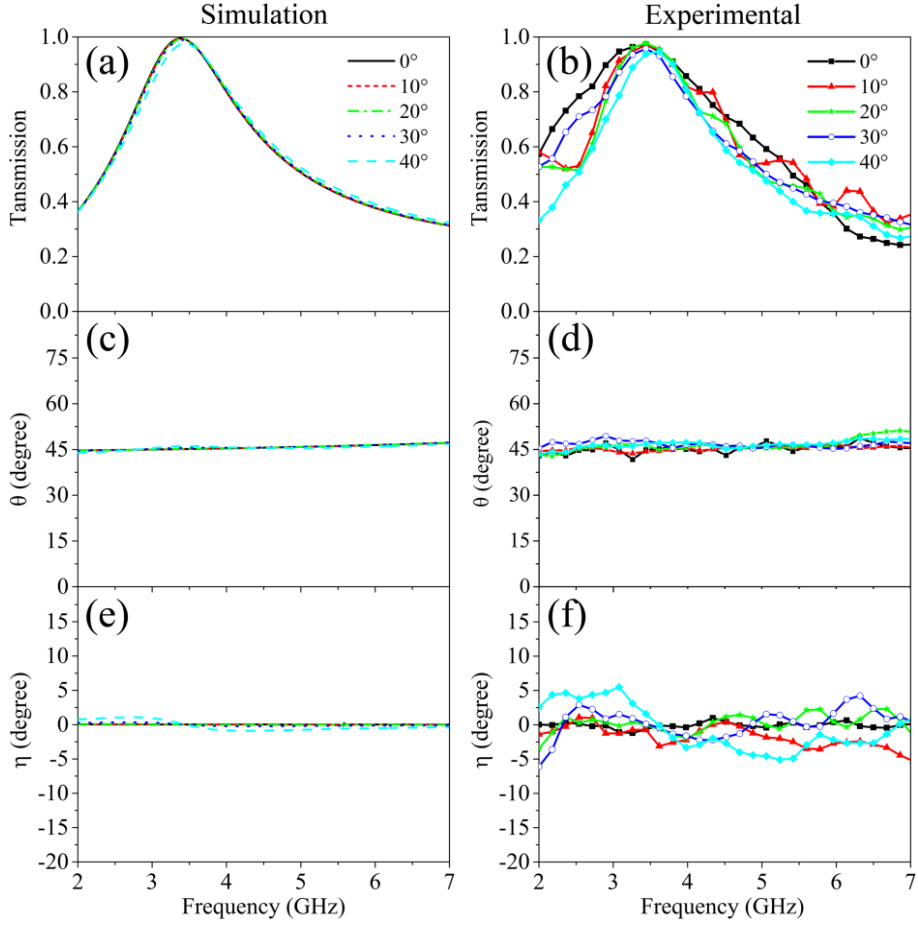

**Figure 3.** The simulation (left column) and experimental (right column) results of the single-layer CMM at oblique incidence. (**a**) and (**b**) Transmission spectra, (**c**) and (**d**) Polarization azimuth rotation angle $\theta$, (**e**) and (**f**) Ellipticity $\eta$.

Generally, the electromagnetic manifestations of metamaterials can be profoundly affected by the permittivity of dielectric substrates. In Fig. 4, further simulations are carried out to investigate the influence of the permittivity of dielectric substrate on the transmission and optical activity of the single-layer CMM. Figure 4(a) portrays the transmission spectrum of the CMM evolving with different permittivity. The results show that, with the permittivity of dielectric substrate increasing, the transmission peak occurs significant red shift and the relative transmission bandwidth with transmittance over 0.8 gradually decreases from 50.3% to 32%, 29.5% and 26.4%. Obviously, in spite of the decrement of relative transmission bandwidth, the CMM can still accomplish broadband and high-performance transmission when the permittivity increases. The red shift of transmission peak and the decrement of transmission bandwidth are attributed to that, the variation of permittivity of dielectric substrate leads to the change of effective permittivity and permeability of the CMM, which further results in the alteration of the effective refractive index and impedance (see Figs S1-S4 in Supplementary). Finally, the transmission spectrum will generate significant change according to equation (11).

The effect of permittivity of substrate on the optical activity of the single-layer CMM is shown in Fig. 4(b). It is noteworthy that the polarization azimuth rotation angle of the CMM has been kept about 45° over all frequency range, without being influenced by the variation of permittivity. This phenomenon can be well explained by equation (12), from which we can see that the polarization azimuth rotation angle is unrelated to the permittivity of dielectric substrate. The aforementioned results confirm that the CMM can realize broadband nondispersive optical activity with high transmission independent of permittivity of dielectric substrate. And these intriguing properties also imply that the CMM might be a good candidate for designing frequency-tunable polarization rotator.

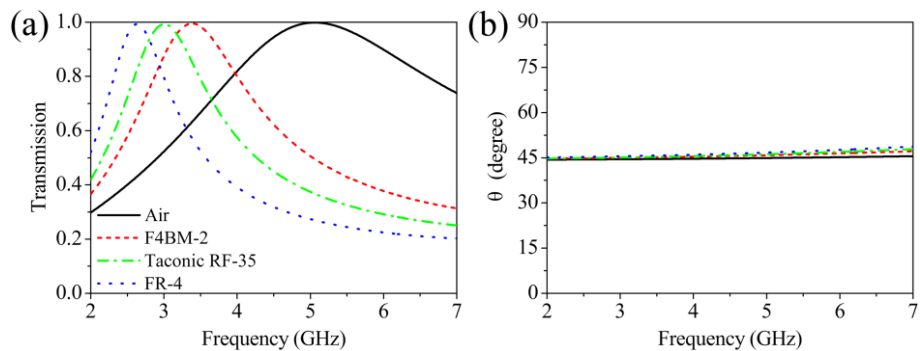

**Figure 4.** The simulated (**a**) transmission spectra and (**b**) polarization azimuth rotation angle of the single-layer CMM with different dielectric substrates. The permittivity of air, F4BM-2, Taconic RF-35, and FR-4 substrates are 1.0+0.00*$i$, 2.65+0.001*$i$, 3.5+0.001*$i$, and 4.6+0.001*$i$, respectively.

**Results of dual-layer CMMs.** The numerical and measured results of the dual-layer CMMs with different interlayer spacings at normal incidence are shown in Fig. 5. It is seen in Fig. 5(a,b) that there are two transmission peaks occurring on the transmission spectra for the dual-layer CMMs. As the interlayer spacing increases, the two transmission peaks exhibit significant red shifts and become close to each other. These phenomena arise due to that, the double layers of CMMs can form a Fabry-Perot-like resonant cavity, and the Fabry-Perot-like resonance will occur when the electromagnetic waves pass through the double layers of CMMs, resulting in the generation of two transmission peaks. Moreover, the Fabry-Perot-like resonant cavity with different space distances will generate different resonant responses, which leads to the transmission spectra altering with the interlayer spacings[48].

Figure 5(c-f) portray the polarization azimuth rotation angle and ellipticity of dual-layer CMMs evolving with different interlayer spacings, respectively. It is obvious that the polarization azimuth rotation angle and ellipticity within the whole frequency region has been kept approximatively 90 ° and 0 °, respectively, and the alteration of the interlayer spacing has almost no influence on the optical activity and ellipticity of the dual-layer CMMs. The fascinating properties mentioned above imply that the dual-layer CMMs can function as a 90 ° polarization rotator, of which the transmission spectrum can be dynamically tuned by varying the space distance between the two layers of CMMs. Furthermore, compared Fig. 5(c,d) with Fig. 2(c,d), it can be found that the optical activity of dual-layer CMMs is just two times of that of the single-layer CMM. And our further simulations demonstrate that the optical activity of multi-layer CMMs is proportion to the number of CMM layers (see Fig. S5 in Supplementary).

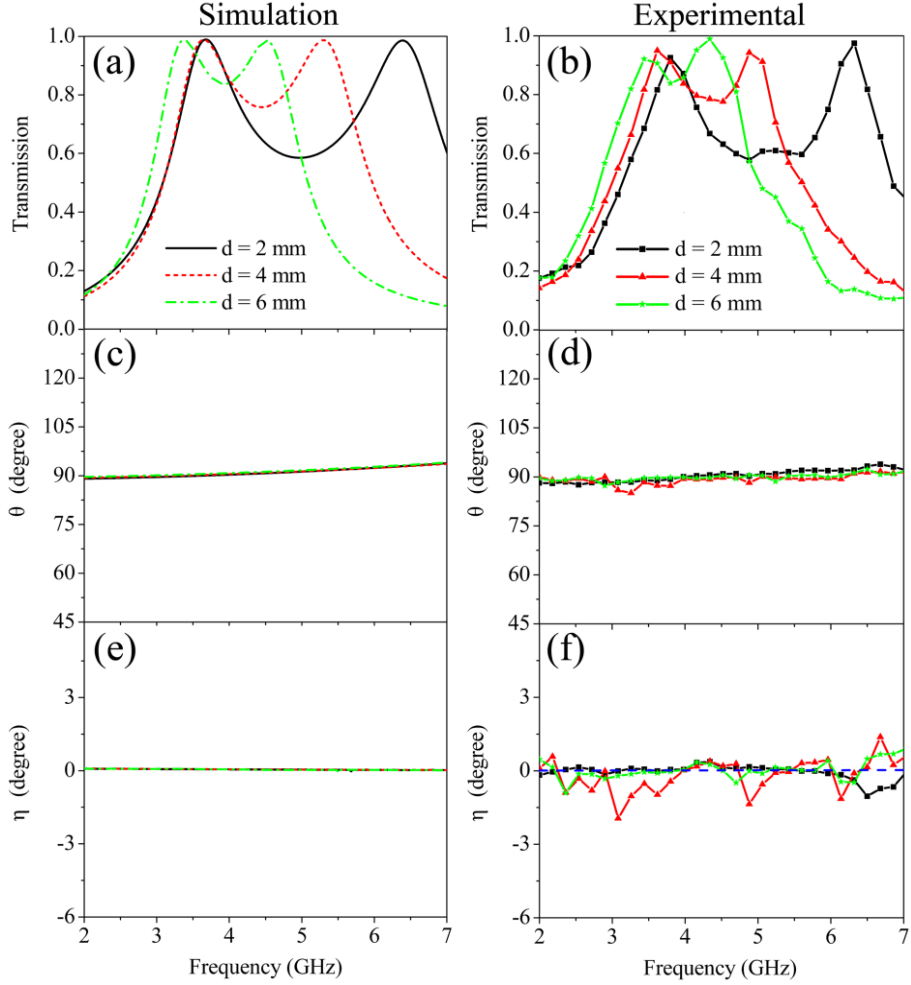

**Figure 5.** The numerical (left column) and measured (right column) results of the dual-layer CMM with different interlayer spacings at normal incidence. (**a**) and (**b**) Transmission spectra, (**c**) and (**d**) Polarization azimuth rotation angle $\theta$, (**e**) and (**f**) Ellipticity $\eta$.

## Discussion

In order to get physical insight into the mechanism of broadband nondispersive optical activity and zero ellipticity, we examine the surface current distributions of the considered CMM, as shown in Fig. 6(a). It is significant that the current flows freely in the interconnected metal patterns. Consequently, the CMM exhibits a nonresonant Drude-like response, which can effectively avoid the dispersive optical activity due to the Lorenz-like resonances existing in the previous CMMs[16,19-21]. Figure 6(b) presents the schematic illustration of the flowing direction of surface current. It is seen that the free-flowing current can induce the magnetic moments **M** being in the opposite direction to the external electric field **E**. And in Fig. 6(c) the simulated magnetic field distribution in the middle plane of the unit cell also confirms the generation of the induced magnetic moments **M**. These results imply that strong cross-coupling effect occurs as the electromagnetic waves pass through the CMM. This is the origin of the chirality of the proposed CMM. Figure 6(d) plots the numerical results of Re($\kappa$), Im($\kappa$), and Re($\kappa$)*$\omega$/2π. Our calculated result confirms that the dissipation constant $\alpha$ in equation (10) is far smaller than the angular frequency $\omega$. As the chirality parameter $\kappa$ is a finite value, the imaginary part of $\kappa$ is thus nearly zero and can be ignored. Then, the equation (10) can be approximatively expressed as $\kappa \approx \pm \frac{\mu_0 caS}{\omega LV} \approx \text{Re}(\kappa)$, from which we can found that the real part of $\kappa$ is inversely proportional to $\omega$. As a result, the value of $\text{Re}(\kappa)*\omega$ is a constant. Finally, the polarization azimuth rotation angle $\theta$ will be constant in the whole frequency range according to equation (12), *i.e.*, the nondispersive optical activity occurs. Additionally, the circular dichroism of CMMs is closely related to the imaginary part of $\kappa$[48]. Since imaginary part of $\kappa$ is approximatively zero, the circular dichroism is therefore absent, which results in the ellipticity being zero.

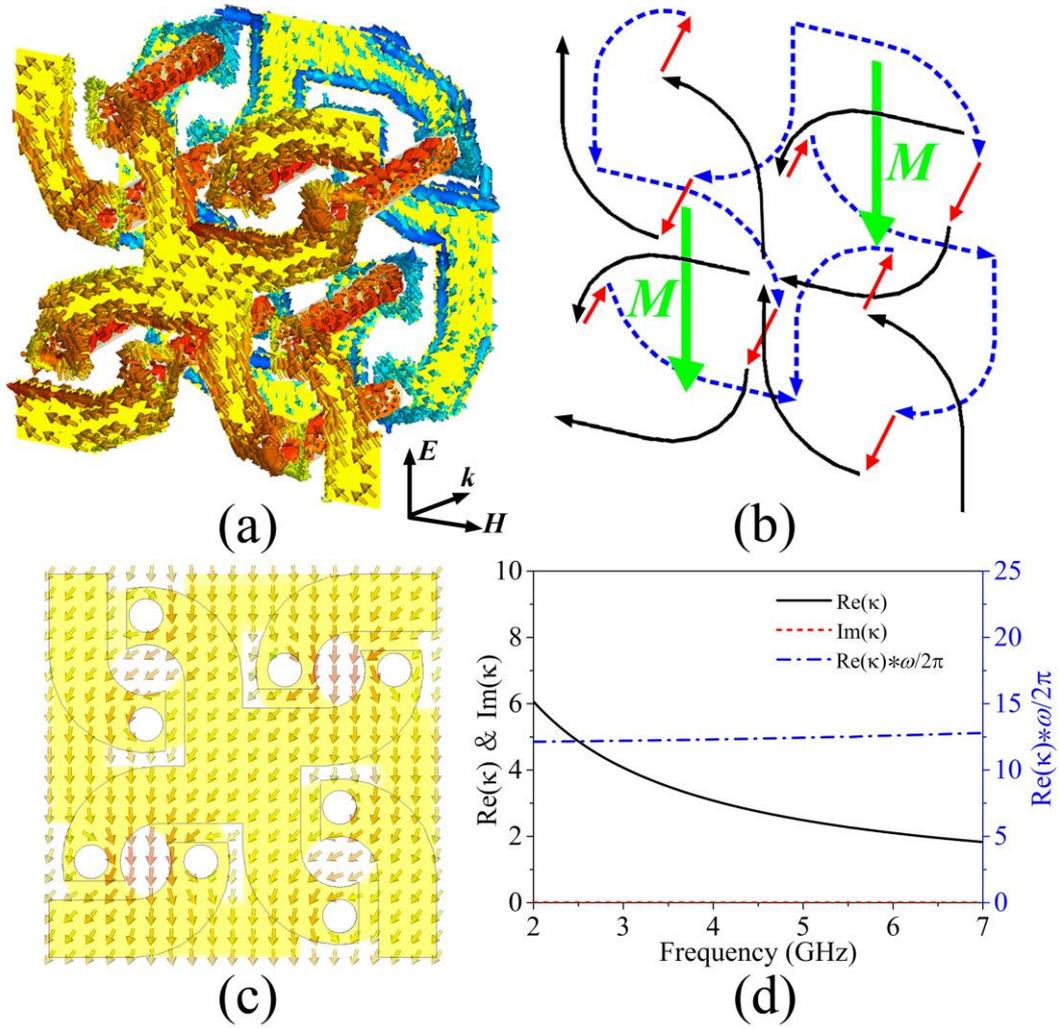

**Figure 6.** (**a**) Surface current distribution of the designed CMM. (**b**) Schematic illustration of the flowing direction of surface current. (**c**) The simulated *H* field distribution at the middle plane of the unit cell. (**d**) The numerically calculated results of Re($\kappa$), Im($\kappa$), and Re($\kappa$)*$\omega$/2$\pi$.

## Conclusions

In summary, we have demonstrated a fascinating CMM that consists of double helical structures. As a result of the interconnected metal structures, this kind of CMM generates Drude-like response, which combines with the C4 symmetry geometry giving rise to the broadband nondispersive optical activity and zero ellipticity simultaneously accompanied by high transmittivity. Most notably, the broadband behaviors of nondispersive optical activity and high transmission of the single-layer CMM are independent of the incident angles of electromagnetic waves and permittivity of dielectric substrate. And the transmission spectrum of the CMM can be successively tuned by varying the permittivity of substrate but with the optical activity unchanged. In addition, the optical activity of multi-layer CMMs is proportional to the number of CMM layers without being affected by the interlayer coupling effect, which enables us to obtain much stronger optical activity just by simply increasing the layer number. With the intriguing properties, the elaborate CMM exhibits more application flexibility and is greatly appealing for controlling the polarization state of electromagnetic waves.

**Methods**

**Numerical simulation.** Simulations were achieved with the commercial software CST Microwave Studio. In the simulations, a linearly polarized wave was incident on the sample; the unit cell boundary conditions were employed in the $x$ and $y$ directions and open boundary conditions were utilized in the $z$ direction.

**Experimental measurement**. The experimental measurements were carried out by using an AV3629 network analyzer with two broadband linearly polarized horn antennas in an anechoic chamber. Using the linear co-polarization and cross-polarization transmission coefficients $T_{xx}$ and $T_{yx}$, we can obtain the transmission coefficients of the circularly polarized waves by the formula $T_{\pm} = T_{xx} \mp i*T_{yx}$, where $T_{+}$ and $T_{-}$ are the transmission coefficients of the right-handed and left-handed circularly polarized waves, respectively. The optical activity is revealed via the polarization azimuth rotation angle $\theta = \frac{\arg(T_{+}) - \arg(T_{-})}{2}$. And the circular dichroism of the transmitted waves is characterized by the ellipticity $\eta = \frac{1}{2}\arctan\frac{|T_+|^2 - |T_-|^2}{|T_+|^2 + |T_-|^2}$.

**Acknowledgements**

This work was supported by the National Natural Science Foundation of China (Grant Nos. 11174234, 11204241, 11404261, 61601375, 61601367), the Fundamental Research Funds for the Central Universities (Grant No. 3102016ZY029, 3102016ZY028) and the Northwestern Polytechnical University Scientific Research Allowance (Grant No. G2015KY0302). The USF portion of this work was supported by the Alfred P. Sloan Research Fellow grant BR2013-123 and by KRISS grant GP2016-034.

**Author Contributions**

K.S. and J.Z. conceived the idea, designed the experiment, and supervised the project. K.S., M.W. and Z.S. performed the numerical simulations. Y.L. and C.D. fabricated the sample and carried out the experimental measurements. K.S., C.L., X.Z., K.B. and J.Z. did the theoretical analysis. K.S. and J.Z. co-wrote the manuscript. All the authors have reviewed the manuscript.